\documentclass[10pt,aps,showpacs,a4paper,floatfix,twocolumn,tightenlines]{revtex4}
\usepackage{epsfig}
\usepackage{psfig}
\usepackage{graphicx}
\usepackage{graphics}
\usepackage{xspace}
\usepackage{amssymb}
\usepackage{amsmath}
\usepackage{latexsym}
\usepackage{natbib}
\usepackage{mathrsfs}
\newcommand{\inieq}{\begin{eqnarray}}            
\newcommand{\fineq}{\end{eqnarray}}            

\def\r{\mbox{\boldmath $r$}}

\def\p{\mbox{\boldmath $p$}}

\def\q{\mbox{\boldmath $q$}}

\def\r{\mbox{\boldmath $r$}}

\begin{document}
\title{ Meson exchange currents in electromagnetic one-nucleon emission }

\author{Carlotta Giusti and Franco Davide Pacati}
\affiliation{Dipartimento di Fisica Nucleare e Teorica, 
Universit\`{a} di Pavia, and \\
Istituto Nazionale di Fisica Nucleare, 
Sezione di Pavia, I-27100 Pavia, Italy}

\date{\today}

\begin{abstract}
The role of meson exchange currents (MEC) in electron- and 
photon-induced one-nucleon emission processes is studied in a nonrelativistic 
model including correlations and final state interactions. The nuclear 
current is the sum of a one-body and of a two-body  part. The two-body current 
includes pion seagull, pion-in-flight and the isobar current contributions. 
Numerical results are presented for the exclusive 
$^{16}$O$\left(e,e^{\prime}p\right)^{15}$N and 
$^{16}$O$\left(\gamma,p\right)^{15}$N reactions. MEC effects are in general 
rather small in $\left(e,e^{\prime}p\right)$, while in $\left(\gamma,p\right)$ 
they are always large and important to obtain a consistent description 
of $\left(e,e^{\prime}p\right)$ and $\left(\gamma,p\right)$ data, with the same 
spectroscopic factors. The calculated $\left(\gamma,p\right)$ cross sections are
sensitive to short-range correlations at high values of the recoil momentum,
where MEC effects are larger and overwhelm the contribution of correlations. 
\end{abstract}
\pacs{25.20.Lj, 25.30.Fj, 24.10.Eq}

\maketitle


\section{introduction}
\label{intro}
Electromagnetic knockout reactions with one-nucleon emission have provided in
the last decades a wealth of information on the single particle (s.p.) properties of
light, medium and heavy nuclei~\cite{FM,PR,Oxford,Kelly1}.
The electromagnetic probe is particularly well-suited for studying the nuclear
properties as its relatively weak interaction with the nuclear matter
allows to penetrate deeply in the nucleus interior and to explore also the inner
states.

The Distorted Wave Impulse Approximation (DWIA) has successfully been applied to
the analysis of experimental data produced by these reactions. 
The overlap integral between the target bound state and
the different states of the residual nucleus, which can be obtained from a
one-body potential able to reproduce the binding energy and the density of the
s.p. states, was found in agreement with the many-body mean-field
calculations.
Moreover, an optical potential derived from elastic nucleon scattering off
nuclei was able to well reproduce the shape of the reduced cross sections, which
are essentially the distorted momentum distributions of nucleons in the nucleus.
This means that the gross features of the reaction are well understood.

Many particular aspects of the reaction mechanism have been studied, like
relativistic corrections of the nuclear current and off-shell effects due to the
binding of the initial nucleon in the nuclear medium.
Moreover, the Coulomb distortion of the incoming and outgoing electron waves in
the electron field of the nucleus was studied~\cite{DWEEPY} and also an exact 
treatment was applied, which calculates the solutions of the
Dirac equations for the partial waves~\cite{Jin,Udias}. 
More recently, complete relativistic calculations~\cite{Ud1,Meucci} 
have been compared with
the new data at higher energy obtained at TJNAF~\cite{Gao}.

Despite these interesting results, some problems are still open and have found
only a partial solution. The low value of the spectroscopic factors, which are
about 60-70 \% of the values predicted by the independent-particle shell model, 
addresses to more complicate mechanisms than the
one-body one in the frame of a mean-field theory. Moreover, consistency
between electron and the photon-induced reactions is required and should 
be obtained in the same theory frame. Strong difficulties, on the contrary, have
to be faced when treating, in particular, the reactions with neutron emission. 
In addition, the study of higher energy processes has shown up ambiguities 
in the choice of the nuclear current operator~\cite{Meucci}, which must be 
eliminated in a consistent theory.

Different mechanisms have been advocated in order to solve these problems.
First of all, nucleon-nucleon correlations, which produce the defect in the 
spectroscopic factors of the hole spectral functions. Theoretical investigations
with different correlation methods~\cite{FC,Mok,Vn97,Po96,Po97,Gai2000} 
indicate that only a few percent of the defect is due to short-range 
correlations (SRC). When tensor correlations are added the depletion amounts to 
$\sim 10$\%, at most $\sim 15$\% in heavy nuclei. Further depletion is given by
long-range correlations~\cite{Ang,Geurts,Amir,BD}. A full and consistent
evaluation  of all the different types of correlations, which affect the 
spectroscopic
factors and the overlap functions to be included in the DWIA approach, is
anyhow still unavailable.

Multi-step processes have also been considered in the treatment of the final 
state interaction of knockout reactions, but they were shown to give a sizable
contribution only at high excitation energy or missing momenta~\cite{Dem}. 

In this paper we discuss the effect of two-body meson exchange currents (MEC) 
both in $\left(e,e^{\prime}p\right)$ and in $\left(\gamma,p\right)$ reactions. 
This topic has already been discussed in some papers with different 
approaches~\cite{BR,Benenti,Sluys,Rick,ALC,Co}. Some disagreements are present 
in the results of the various groups. In general, the effects were found 
significant, but large only in the region of high recoil momenta. Moreover, in 
the $\left(e,e^{\prime}p\right)$ reaction the response functions appear more
sensitive to MEC effects than the total cross sections. It is not surprising 
that some differences can be found between the results of the different models, 
even when the operators describing MEC are identical, due to interference 
between the different ingredients of the calculations. A further study seems 
therefore useful to clarify the situation.

In this work MEC effects are evaluated in the frame of a nonrelativistic direct
knockout (DKO)model with final state interactions. The direct one-body 
contribution corresponds to the DWIA approach,
which was successfully applied to the analysis of $\left(e,e^{\prime}p\right)$ 
data at moderate energy~\cite{Oxford,NIKHEF}. The interaction occurs in the
initial state through a one-body and a two-body current with a pair of nucleons.
Only one nucleon is emitted and the other one is reabsorbed in the residual
nucleus. Correlations can be directly included in the s.p. wave functions or 
with a central correlation function. This approach was originally proposed in 
Ref.~\cite{Benenti} and has more recently been applied to the
$\left(\gamma,p\right)$ reaction~\cite{Gai2000,Iva2001} and, in a relativistic 
model, to the $\left(e,e^{\prime}p\right)$ and $\left(\gamma,p\right)$ 
reactions~\cite{MGP}. In these previous applications, however, only the 
pion-seagull term was included in the two-body current. Moreover, in 
Refs.~\cite{Benenti,Gai2000,Iva2001} a simplified treatment of the spin was 
used in order to reduce the complexity of the numerical calculations, and the 
spin-orbit term was neglected in the optical potential. 

A more refined theoretical model is considered in this paper. The two-body 
current includes the contribution of all the diagrams with one-pion exchange, 
namely seagull, 
pion-in-flight, and the diagrams with intermediate $\Delta$-isobar 
configurations. Moreover, also spin coupling and the spin-orbit term of the 
optical potential are fully included in the calculations. Application of this
model to the $\left(e,e^{\prime}p\right)$ and $\left(\gamma,p\right)$ reactions
and comparison with data allow us to check the consistency of the description 
of electron and photon-induced reactions within the same theoretical framework.

In Sec.~\ref{sec.th} the relevant formalism and the parameters used in the 
calculation are given. Numerical results are presented and discussed in 
Sec.~\ref{sec.results} for the cross section and the structure functions of the
exclusive $^{16}$O$\left(e,e^{\prime}p\right)^{15}$N reaction and for the cross
section of the exclusive $^{16}$O$\left(\gamma,p\right)^{15}$N reaction. 
Some conclusions are drawn in Sec.~\ref{con}.

\section{theoretical model }
\label{sec.th}

The coincidence unpolarized cross section for the reaction induced by an 
electron, with momentum $\p_{0}$ and energy $E_0$, with
$E_0=|\p_0|=p_0$, where a nucleon, with momentum $\p'$ and energy $E^{\prime}$,
is ejected from a nucleus, is given, in the one-photon exchange approximation 
and after integrating over the energy of the emitted nucleon, by~\cite{Oxford}

\begin{eqnarray}
\frac{{\rm d}^{5}\sigma}{{\rm d}E'_{0}{\rm d}\Omega'_{0}{\rm d}\Omega'} = 
&K& [2\epsilon_{\rm L}f_{00}+f_{11}-\epsilon f_{1-1}\cos2\alpha \nonumber \\
   \nonumber \\
    & + & \sqrt{\epsilon_{\rm L}(1+\epsilon)}f_{01}\cos\alpha] . \label{eq:cs}
\end{eqnarray}
\newline
Here $E'_0$ is the energy of the scattered electron, $\alpha$ the angle between
the plane of the electrons and the plane containing the momentum transfer $\q$
and $\p'$. The quantity 

\begin{equation}
\epsilon = \left(1-\frac{2\q^{2}}{q^{2}_{\nu}}\tan^{2}\frac{\theta}{2}\right)
^{-1}
\end{equation}
\newline 
measures the polarization of the virtual photon exchanged by the electron 
scattered at an angle $\theta$ and 

\begin{equation}
\epsilon_{\rm L} = -\frac{q^{2}_{\nu}}{\q^{2}}\epsilon    ,
\end{equation}
\newline 
where $q^{2}_{\mu}=\omega^{2}-\q^{2}$, with $\omega = p_0 - p'_0$ and 
$\q = \p_0 - \p'_0$, is the four-momentum transfer. The coefficient 

\begin{equation}
K = \frac{e^{4}}{16\pi^{2}}\frac{\Omega_{\rm f}} {f_{\rm R}}
\frac{E'_0} {E_0q^{2}_{\nu}(\epsilon-1)}, 
\end{equation}
\newline
contains the phase-space factor $\Omega_{\rm f} = p' E'$ and the recoil
factor $f_{\rm R}^{-1}$, with 

\begin{equation}
 f_{\rm R}= 1 + \frac{E'}{E_{\rm B}}\left( 1 - \frac{q}{p'} \cos \gamma\right),
\label{eq:rec}
\end{equation}
\newline
where $E_{\rm B}$ is the total relativistic energy of the residual nucleus and
$\gamma$ is the angle between $\q$ and $\p'$.

The structure functions $f_{\lambda\lambda'}$ represent the response of the
nucleus to the longitudinal ($\lambda$=0) and transverse ($\lambda=\pm 1$)
components of the electromagnetic interaction and only depend on $\omega,q,p'$
and $\gamma$.

When the nucleon is emitted by a real unpolarized photon, only $f_{11}$
contributes. Then the cross section of the $(\gamma,N)$ reaction reads

\begin{equation}
\frac{{\rm d}^{2}\sigma}{{\rm d}\Omega'} = \frac{\pi e^2}{2 E_\gamma}
\frac{\Omega_{\rm f}}{f_{\rm R}} f_{11}. 
\label{eq:csg}
\end{equation}
\newline
where $E_\gamma$ is the energy of the incident photon. 

The structure functions are given in terms of the hadron tensor~\cite{Oxford}

\begin{equation} 
W^{\mu \nu} = \overline{\sum_{\mathrm{i}}}
{\sum_{\mathrm{f}}} J^\mu (\q) J^{\nu *} (\q) 
\delta (E_{\mathrm{i}} - E_{\mathrm{f}}) ,
\label{eq:htens}
\end{equation}
\newline
The quantities $J^\mu(\q)$ are the Fourier transforms of the transition
matrix elements of the nuclear charge-current density operator between initial
and final nuclear states
\begin{equation}
J^{\mu}({\mbox{\boldmath $q$}}) = \int \langle\Psi_{\rm{f}}
|\hat{J}^{\mu}({\mbox{\boldmath $r$}})|\Psi_{\rm{i}}\rangle
{\rm{e}}^{\,{\rm{i}}
{\footnotesize {\mbox{\boldmath $q$}}}
\cdot
{\footnotesize {\mbox{\boldmath $r$}}}
} {\rm d}{\mbox{\boldmath $r$}} .     \label{eq:jm}
\end{equation}

The nuclear current operator is given in terms of one-body and
two-body components. It can be written in the 
coordinate representation as

\begin{equation}
\hat{J}^{\mu}(\r)=\sum_{i=1}^A J^{(1)\mu}(\r,\r_i)+
\sum_{i< j=1}^A J^{(2)\mu}(\r,\r_i,\r_j), 
\label{eq:no}
\end{equation}
\newline
where the spin and isospin indices are omitted for simplicity.

Taking into account the antisymmetrization of the initial and final nuclear
state and including a correlation function of Jastrow type $f(r_{ij})$, 
the current matrix
element can be written, under the assumption that the interaction occurs only 
on a couple of correlated nucleons, as 

\inieq
\langle \Psi_{\textrm {f}}\mid J^{\mu}\mid \Psi_{\textrm {i}}\rangle \simeq 
\qquad \qquad \qquad \qquad \qquad \qquad 
\nonumber \\
\nonumber \\
\sum_{\alpha = 1}^{\textrm{A}} \langle \chi^{(-)}(\r_1)
\varphi_{\alpha}(\r_2)\mid j^{\mu}(\r,\r_1,\r_2)\mid 
f(r_{12})
\nonumber \\ 
\nonumber \\
\times  (\varphi_{\beta}(\r_1)\varphi_{\alpha}(\r_2)
-\varphi_{\alpha}(\r_1)\varphi_{\beta}(\r_2))\rangle 
\ , \label{eq:cca}
\end{eqnarray}
\newline
where where $\chi^{(-)}$ is the distorted wave function of the outgoing nucleon 
and $\varphi_{\alpha(\beta)}$ are s.p. shell-model wave functions.

The current operator can be written as
 
\inieq
j^{\mu}(\r,\r_1,\r_2) & =& \frac{1}{A-1}\left[J^{(1)\mu}(\r,\r_1)+
J^{(1)\mu}(\r,\r_2)\right] \nonumber \\
\nonumber \\
& + & J^{(2)\mu}(\r,\r_1,\r_2).
\label{eq:nc}
\end{eqnarray}
\newline

Therefore, in the model the interaction occurs through one-body and two-body 
currents with a pair of correlated nucleons. Only one nucleon is emitted and 
the other nucleon is reabsorbed in the residual nucleus. For the nucleon which 
is not emitted a sum over all the the s.p. states is included in  
Eq.~(\ref{eq:cca}).
The one-body current contains a direct and an exchange term. When the
correlation function $f(r_{ij})$ is not included in Eq.~(\ref{eq:cca}) only the
direct term survives in the one-body current: its contribution corresponds to
the DKO mechanism and gives the DWIA result.

In the calculations the one-body current includes the convective and the spin
terms. The two-body current is derived by performing a non relativistic 
reduction of the lowest-order Feynman diagrams with one-pion exchange.
We have thus currents 
corresponding to the seagull and pion-in-flight diagrams and to the diagrams 
with intermediate isobar configurations.

The operator form of the $\Delta$ current has been derived in 
Ref.~\cite{WAGP}. It is given by the sum of the contributions of two types 
of processes, corresponding to the $\Delta$-excitation and 
$\Delta$-deexcitation currents. The first process (I) describes 
$\Delta$-excitation by photon absorption and subsequent deexcitation by pion 
exchange, while the second process (II) describes the time interchange of the 
two steps, i.e., first excitation of a virtual $\Delta$ by pion exchange in a 
NN collision and subsequent deexcitation by photon absorption. The propagator 
of the resonance, $G_\Delta$, depends on the invariant energy $\sqrt s$ of the
$\Delta$, which is different for processes I and II. For the deexcitation 
current the static approximation can be applied, i.e. 
\begin{equation}
G_\Delta^{\,\mathrm{II}} = (M_\Delta-M)^{-1},    \label{eq:ei}
\end{equation}
where $M_{\Delta} = 1232$ MeV. For the excitation current we use~\cite{WWA}
\begin{equation}
G_\Delta^{\,\mathrm{I}} = \left({M_\Delta}-\sqrt{s_{\mathrm I}}-
\frac{{\mathrm{i}}}{2}\Gamma_\Delta (\sqrt{s_{\mathrm I}}) \right)^{-1},
\label{eq:prop}
\end{equation}
with 
\begin{equation}
\sqrt{s_{\mathrm I}}=\sqrt{s_{NN}}-M,
\label{eq:s1}
\end{equation}
where $\sqrt{s_{NN}}$ is the invariant energy
of the two interacting nucleons. The energy-dependent decay width of the 
$\Delta$, $\Gamma_\Delta$, has been taken in the calculations according to the 
parameterization of Ref.~\cite{BM}.
 
The operators of the two-body current have been corrected for their behaviour at
short distances with a hadronic form factor, which was chosen as a monopole
function with a cut-off parameter $\Lambda_\pi$ = 800 MeV.

\section{results}
\label{sec.results}

Calculations have been performed for exclusive 
$\left(e,e^{\prime}p\right)$ and $\left(\gamma,p\right)$ reactions from
$^{16}$O. Although the model can in principle be applied to any other target 
nucleus, we do not expect, on the basis of previous investigations, different 
effects of two-body currents for different nuclei. The large amount of 
theoretical and experimental work carried out on $^{16}$O, an ample choice of 
theoretical ingredients available for our calculations, and the possibility of 
a direct comparison with different sets of data make $^{16}$O a well suited 
target for our study, which allows us a comparison with our previous 
calculations and with the results of other and different theoretical models. 
Moreover, calculations on $^{16}$O make it possible to check the consistency 
of the description of $\left(e,e^{\prime}p\right)$ and $\left(\gamma,p\right)$ 
reactions in comparison with data.

\subsection{The \bf{$^{16}$O$\left(e,e^{\prime}p\right)^{15}$N} reaction}


\begin{figure}
\includegraphics[height=7cm, width=8.4cm]{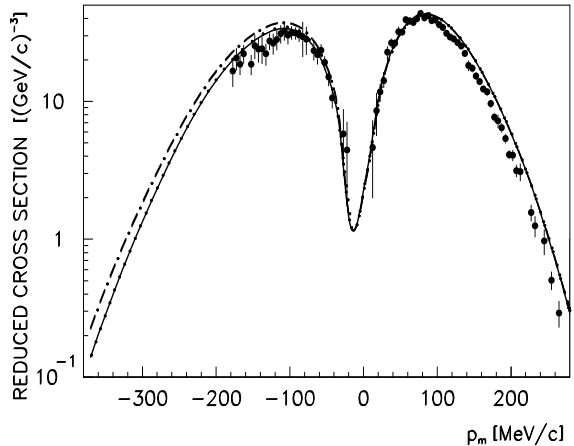} 
\caption {The reduced cross section of the 
$^{16}$O$\left(e,e^{\prime}p\right)^{15}$N$_{\mathrm{g.s.}}$ reaction  as a 
function of the recoil momentum $p_{\mathrm{m}}$ in parallel kinematics, with 
$E_0 =$ 520 MeV and an outgoing proton energy 
$T_{\mathrm p} =$ 90 MeV. The data are from~\cite {Leus}. The optical 
potential is taken from~\cite{Schwandt} and the overlap function is derived
from the OBDM of~\cite{Po96}. The dotted line gives the contribution of 
the one-body current (DWIA). The other lines have been obtained by adding the 
various terms of the two-body current: one-body + seagull (dot-dashed lines), 
one-body + seagull + pion-in-flight (dashed lines), one-body+ seagull+ 
pion-in-flight+ $\Delta$ (solid lines). A reduction factor 0.8 has been 
applied to the theoretical results. Positive (negative) values of 
$p_{\mathrm m}$ refer to situations where $|\q|<|\p'|$ ($|\q|>|\p'|$).}
\label{fig:fig1}
\end{figure}

A numerical example is shown in Fig.~\ref{fig:fig1} for the 
$^{16}$O$\left(e,e^{\prime}p\right)$ reaction and the transition to the 
$1/2^{-}$ ground state of $^{15}$N in the parallel kinematics of the experiment 
carried out at NIKHEF~\cite{Leus}. In parallel kinematics the momentum of the 
outgoing nucleon $\p^{\prime}$ is fixed and is taken parallel or antiparallel 
to the momentum transfer $\q$. Different values of the recoil momentum  
$\p_{\mathrm m}$ are obtained by varying the electron scattering angle and 
therefore the magnitude of the momentum transfer. In order to allow a comparison
with data, the results are presented in terms of the reduced cross 
section~\cite{Oxford}, which is defined as the cross section divided by a 
kinematical factor and the elementary off-shell electron-proton scattering 
cross section, usually taken on the base of the CC1 prescription~\cite{deF}. 
Final state interactions are taken into account by the same phenomenological 
optical potential~\cite{Schwandt} as in the original DWIA analysis of 
data~\cite{Leus}. 
The Coulomb distortion of electron waves is included through the effective 
momentum approximation~\cite{Oxford,DWEEPY}, which is a good approximation for 
a nucleus as light as $^{16}$O. 

For the bound state wave function of the emitted proton we have used the 
one-nucleon overlap function extracted in Ref.~\cite{Gai2000} from the 
asymptotic behaviour of the one-body density matrix (OBDM) of Ref.~\cite{Po96}, 
which in the analysis of Ref.~\cite{Gai2000} was able to give the best and a 
consistent description of $^{16}$O$\left(e,e^{\prime}p\right)^{15}$N and 
$^{16}$O$\left(\gamma,p\right)^{15}$N data. The s.p. wave functions of the
second nucleon in the sum of Eq.~(\ref{eq:cca}) are consistently evaluated for 
all the occupied proton and neutron states.

The overlap function already includes SRC and tensor correlations 
and a spectroscopic factor (0.9). In order to reproduce the size of the 
experimental data, a reduction factor (0.8) was applied in Ref.~\cite{Gai2000} 
to the calculated reduced cross section. This factor, which is 
applied also in Fig.~\ref{fig:fig1}, can be considered as a further 
spectroscopic factor to be mostly ascribed to the correlations not included in 
the OBDM, namely long-range correlations, which also cause a 
depletion of the quasihole states. In order to avoid possible double counting, 
we do not include a correlation function in the two-nucleon wave function of 
Eq.~(\ref{eq:cca}). Thus, the result with the one-body current in 
Fig.~\ref{fig:fig1} is due only to the direct contribution and in practice 
corresponds to the DWIA result of Ref.~\cite{Gai2000}. In any case, no 
significant effect would be produced by a correlation function in the 
kinematics here considered for the $\left(e,e^{\prime}p\right)$ reaction. 

A different choice of the overlap function would produce a different result, 
but would not change the role of two-body currents, which is practically 
negligible in Fig.~\ref{fig:fig1}. The seagull current produces a slight 
enhancement of the reduced cross section calculated with only the one-body 
current. This enhancement is more visible in the left side of the figure, where 
higher values of the momentum transfer are involved. No significant effect is 
obtained when the pion-in-flight term is added, while the $\Delta$-isobar 
current reduces the calculated cross section and practically cancels the 
contribution of MEC. This result confirms that a DWIA approach with only a 
one-body current is able to give a good description of 
$\left(e,e^{\prime}p\right)$ cross sections.

\begin{figure}
\includegraphics[height=14cm, width=8.4cm]{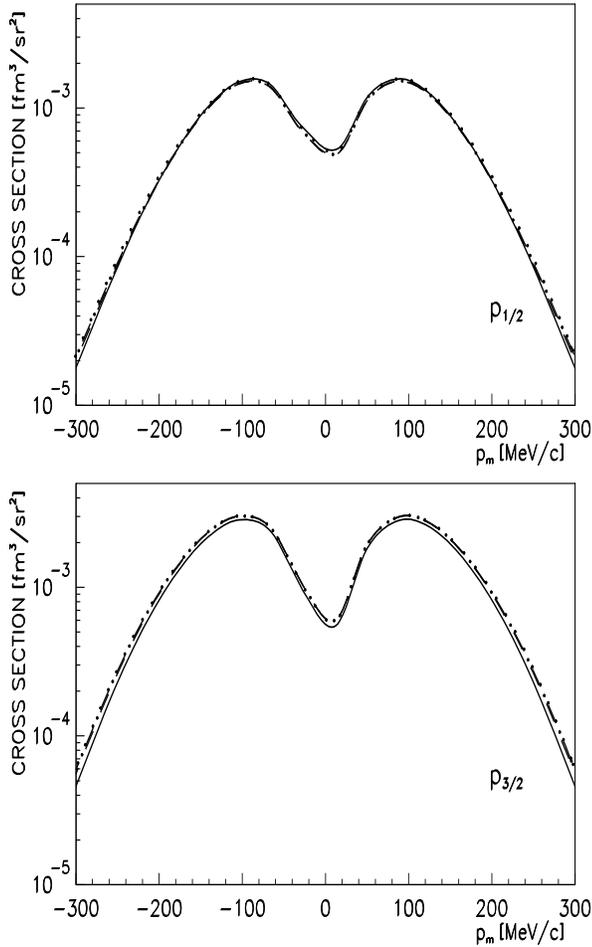} 
\caption {The cross section  of the 
$^{16}$O$\left(e,e^{\prime}p\right)$ reaction  as a function of the recoil 
momentum $p_{\mathrm{m}}$ for the transitions to the $1/2^{-}$ ground state and 
to the $3/2^{-}$ excited state of $^{15}$N in a kinematics with constant 
($\q,\omega$), with $E_{0} = 2000$ MeV and $T_{\mathrm p} = 100$ MeV. 
The optical potential is taken from~\cite{Schwandt}. Overlap functions and line 
convention as in Fig.~\ref{fig:fig1}. Positive (negative) values of 
$p_{\mathrm m}$ refer to situations where the angle between $\p'$ and the 
incident electron $\p_0$ is larger (smaller) than the angle between $\q$ and 
$\p_0$}
\label{fig:fig2}
\end{figure}

Another example is presented in Fig.~\ref{fig:fig2}, where the cross section of 
the $^{16}$O$\left(e,e^{\prime}p\right)$ reaction is displayed for the 
transitions to the $1/2^{-}$ ground state and to the $3/2^{-}$ excited state of 
$^{15}$N in a kinematics where the energy and momentum transfer are constant, 
the outgoing proton energy is fixed, and different values of 
$p_{\mathrm{m}}$ are obtained changing the scattering angle of the outgoing
proton. Also in this case MEC effects are very small. For 
the ground state all the curves practically overlap. For the $3/2^{-}$ state 
only a slight reduction of the calculated cross section is produced by the
$\Delta$ current over the whole distribution.

\begin{figure}
\includegraphics[height=15cm, width=8.4cm]{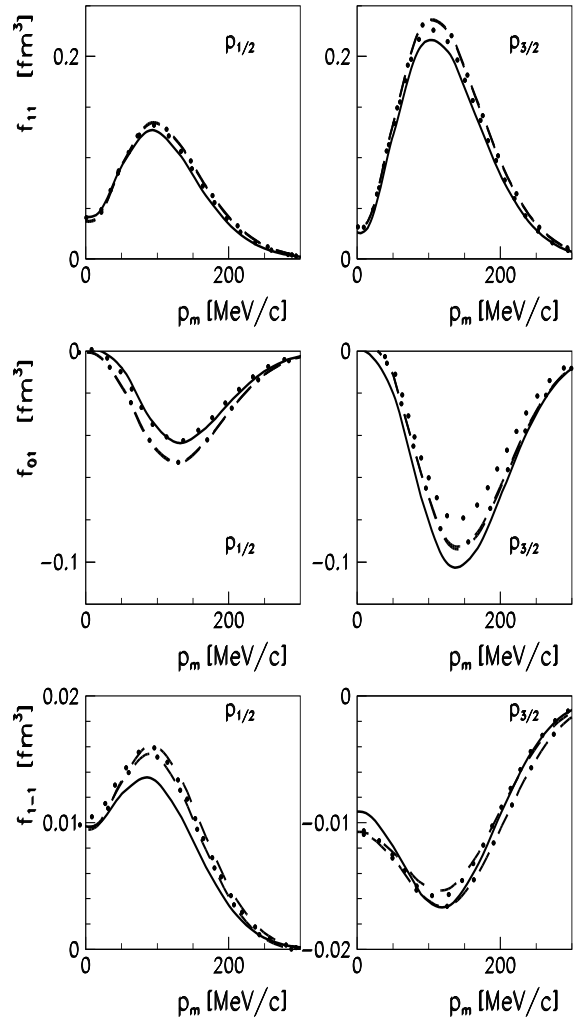} 
\caption {The structure functions, $f_{11}$, $f_{01}$, and $f_{1-1}$, of the 
$^{16}$O$\left(e,e^{\prime}p\right)$ reaction  as a function of the recoil 
momentum $p_{\mathrm{m}}$ for the transitions to the $1/2^{-}$ ground state and 
to the $3/2^{-}$ excited state of $^{15}$N in the same kinematics as in 
Fig.~\ref{fig:fig2}. Optical potential, overlap functions and line convention 
as in Fig.~\ref{fig:fig2}.}
\label{fig:fig3}
\end{figure}

The effects of the two-body currents on the structure functions $f_{11}$, 
$f_{01}$ and $f_{1-1}$ are shown in Fig.~\ref{fig:fig3} in the same conditions
and kinematics as in Fig.~\ref{fig:fig2}. The longitudinal component $f_{00}$, 
which is not affected by MEC in our model, is not presented in 
the figure. MEC effects on $f_{11}$ are similar to the results found for the 
cross section. The seagull term gives a slight enhancement of the one-body
result, while the $\Delta$ current gives a reduction. The total contribution
gives a slight reduction of $f_{11}$ for both transitions. 
For $f_{1-1}$ a small reduction is obtained for the $1/2^{-}$ state and 
negligible effects  for the $3/2^{-}$ state. This structure function is however 
very small. The effects of MEC on the interference function $f_{01}$ are 
different for the two final states. For the ground state the $\Delta$ current 
cancels the effect produced by the seagull term and makes the contribution of 
the two-body currents negligible. For the $3/2^{-}$ state 
the seagull and $\Delta$ currents contributions sum up in the final 
result and produce a sizable effect on $f_{01}$. 

A different effect for the two transitions, with a larger contribution of MEC 
in the $3/2^{-}$ state, was found also in Ref.~\cite{Sluys}. The effects of the 
isobar currents obtained in our work, however, are much smaller. 
Our results also differ from those of Ref.~\cite{BR}, where a 
large contribution was given by the $\Delta$ current on $f_{11}$ and $f_{1-1}$ 
and also on the cross section, but no significant effect was found on $f_{01}$. 

Our results indicate that MEC effects on the exclusive 
$^{16}$O$\left(e,e^{\prime}p\right)^{15}$N are in general rather small and of
about the same order and relevance as in the analysis of Ref.~\cite{ALC}, where, 
however, no particular difference was found for the structure
function $f_{01}$ in the $1 /2^{-}$ and $3/2^{-}$ states.

Very small MEC effects on the $^{16}$O$\left(e,e^{\prime}p\right)^{15}$N 
reaction are found also in the relativistic analysis of Ref.~\cite{MGP}, where 
the calculation of MEC is implemented within the same theoretical framework, 
but only the seagull term is included in the two-body current.

\subsection{The \bf{$^{16}$O$\left(\gamma,p\right)^{15}$N} reaction}


\begin{figure}
\includegraphics[height=10cm, width=8.4cm]{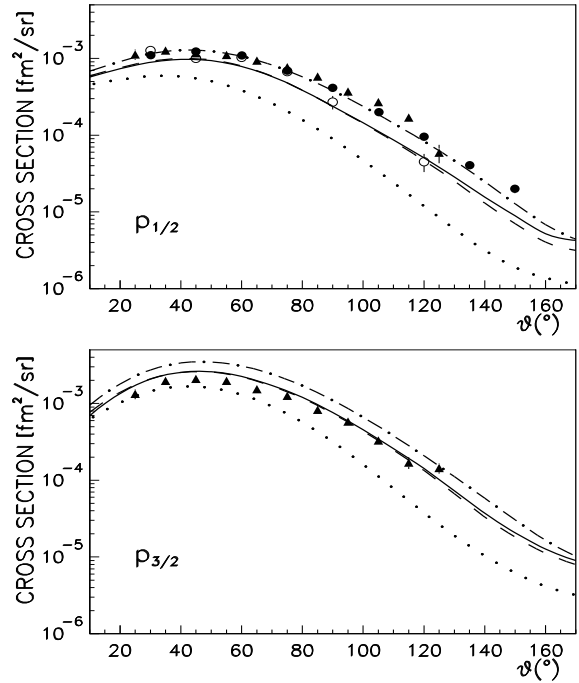} 
\caption {The cross section of the
$^{16}$O$\left(\gamma,p\right)^{15}$N$_{\mathrm{g.s.}}$ reaction as a function 
of the proton 
scattering angle at $E_{\gamma} = 60$ MeV for the transitions to the $1/2^{-}$ 
ground state and to the $3/2^{-}$ excited state of $^{15}$N. The optical 
potential is taken from~\cite{Schwandt} and the overlap functions are derived
from the OBDM of~\cite{Po96}. Line convention as in Fig.~\ref{fig:fig1}. The 
experimental data are taken from~\cite{FO} (black circles),~\cite{Smet} (open 
circles) and~\cite{Miller} (triangles). The theoretical results have been 
multiplied by the reduction factors extracted in~\cite{Gai2000} in comparison 
with  $\left(e,e^{\prime}p\right)$ data, that is 0.8 for $1/2^{-}$ and 0.62 for 
$3/2^{-}$. }
\label{fig:fig4}
\end{figure}
The angular distributions of the $^{16}$O$\left(\gamma,p\right)$ reaction 
at $E_{\gamma} = 60$ MeV for the transitions to the $1/2^{-}$ ground state and 
to the $3/2^{-}$ excited state of $^{15}$N are shown in Fig.~\ref{fig:fig4}. In 
order to check the consistency of $\left(e,e^{\prime}p\right)$ and 
$\left(\gamma,p\right)$ results in comparison with data and allow a direct 
comparison with our previous calculations~\cite{Gai2000}, the same theoretical 
ingredients, i.e. overlap functions and consistent optical potentials, have 
been adopted as in Ref.~\cite{Gai2000} and in Fig.~\ref{fig:fig1}. Moreover, 
the reduction factors obtained in comparison with the 
$\left(e,e^{\prime}p\right)$ data have been applied to the calculated cross 
sections. 

In Fig.~\ref{fig:fig4} the contribution of the one-body current represents a 
large part of the measured cross section at low scattering angles, but is 
unable to describe data. A significant enhancement is produced by the seagull
current and a reasonable description of data is obtained for the transition to 
the ground state. This 
result was already found in Refs.~\cite{Benenti,Gai2000}: the slight numerical 
differences are due to the more 
refined treatment of the spin in the present model, where spin coupling is 
included and the optical potential contains also the spin-orbit term. 

A significant reduction of the cross section is obtained when the pion-in-flight 
term is added, while 
the $\Delta$ current is not very important at the considered value of the 
photon energy. Thus, MEC effects are overestimated by the seagull term, which 
gives the main contribution of MEC at $E_\gamma = 60$ MeV, but is unable to
describe the final effect of two-body currents on the cross section.
As a consequence of the reduction produced by the pion-in-flight current, 
the experimental cross section for the transition to the ground state is 
somewhat underestimated, while a better agreement with data is found for the 
$3/2^{-}$ state. Anyhow, MEC effects remain large and important to 
improve the agreement with data. 


\begin{figure}
\includegraphics[height=10cm, width=8.4cm]{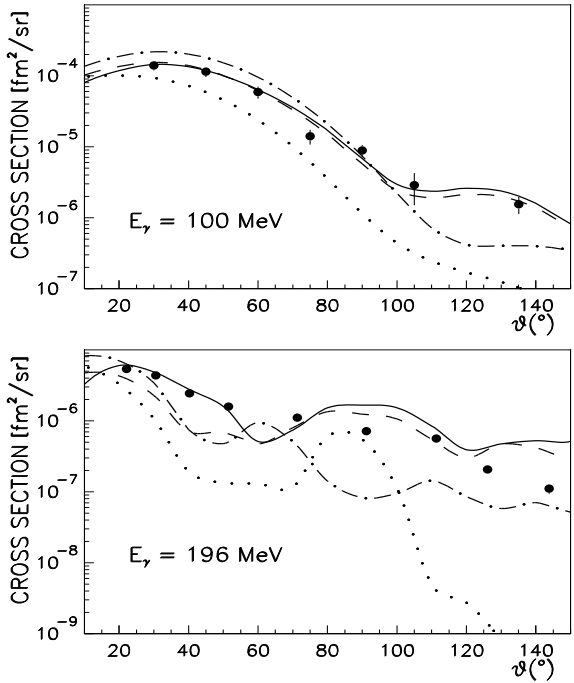} 
\caption {The cross sections of the $^{16}$O$\left(\gamma,p\right)^{15}$N 
reaction as a function of the proton scattering angle at $E_{\gamma} = 100$ and 
196 MeV. Overlap function and line convention as in Fig. 4.  The optical 
potential is taken from~\cite{Schwandt}. The experimental data are taken 
from~\cite{FO} ($E_\gamma = 100$) and~\cite{adams} ($E_\gamma = 196$). 
The theoretical results have been multiplied by the reduction factor 0.8 as in 
Figs. 1 and 4.}
\label{fig:fig5}
\end{figure}
The role of the different terms of the two-body current depends on the photon 
energy. An example is shown in Fig.~\ref{fig:fig5}, where the cross sections of 
the $^{16}$O$\left(\gamma,p\right)^{15}$N$ _{\mathrm{g.s.}}$ reaction at 
$E_{\gamma} = 100$ and 196 MeV are displayed. MEC effects are always large. The
role of the seagull current decreases increasing the photon energy. Important 
effects are given by the pion-in-flight and, at 196 MeV, also by the $\Delta$ 
current. A good agreement with data is achieved at 100 MeV when the 
pion-in-flight term is added. Here pion-in-flight reduces the contribution 
given by seagull for nucleon emission angles up to $\sim 100^{\rm o}$, while at 
larger angles it produces a significant enhancement of the cross section. At 
196 MeV a good agreement with data is obtained for nucleon emission angles up 
to $\sim 70^{\rm o}$ when the $\Delta$ current is added, while for 
larger angles the strong enhancement produced mainly by the pion-in-flight term 
leads to some overestimation of the experimental cross section. In this region, 
however, recoil-momentum values around 700-800 MeV/$c$ are probed, where the 
behaviour of the wave function may become critical and other effects might come
into play. Moreover, the convergence of the integrals in the calculation of
the matrix elements of the two-body current becomes slow. 

Our results indicate that two-body currents give an important contribution to 
$\left(\gamma,p\right)$ cross sections at all the considered photon energies. 
When all the one-pion-exchange diagrams are included in the model, a reasonable 
description of $\left(\gamma,p\right)$ and 
$\left(e,e^{\prime}p\right)$ data is obtained for recoil 
momentum values up to $\sim 500-600$ MeV/$c$ with a consistent choice of the 
theoretical ingredients and with the same spectroscopic factors.  


\begin{figure}
\includegraphics[height=10cm, width=8.4cm]{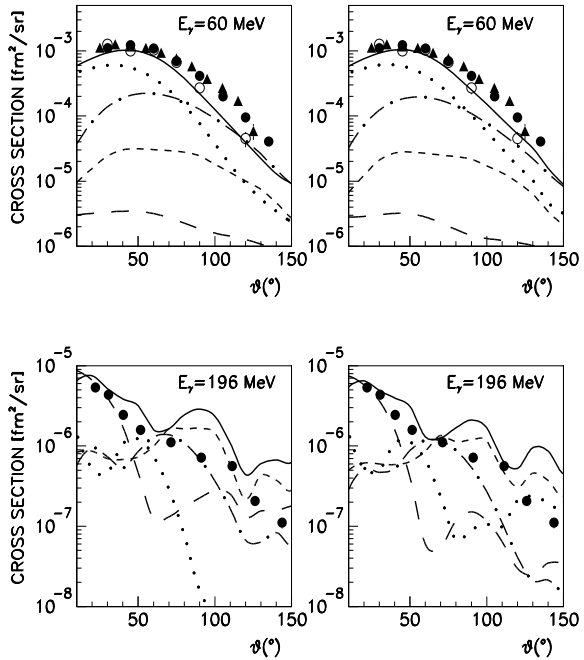} 
\caption {The cross sections of the 
$^{16}$O$\left(\gamma,p\right)^{15}$N$ _{\mathrm{g.s.}}$ 
reaction as a function of the proton scattering angle at $E_{\gamma} = 60$ and 
196 MeV. Optical potential and experimental data as in Figs.~\ref{fig:fig4} and 
\ref{fig:fig5}. The bound state wave functions are taken from~\cite{ES}. The 
theoretical results shown in the right panels have been obtained including in 
the two-nucleon wave function the correlation function of~\cite{GD}. Line 
convention for dotted and solid lines as in Fig.~\ref{fig:fig1}. The 
dot-dashed, long-dashed and short-dashed lines give the pure contributions of 
the seagull, $\Delta$ and pion-in-flight currents, respectively. The 
theoretical results have been multiplied by the reduction factor 0.8, 
consistently with the analysis of $\left(e,e^{\prime}p\right)$ data for the 
same bound state wave function. }
\label{fig:fig6}
\end{figure}

The effect of SRC is investigated in Fig.~\ref{fig:fig6} for the 
$^{16}$O$\left(\gamma,p\right)^{15}$N$ _{\mathrm{g.s.}}$ reaction at 
$E_{\gamma}= 60$ and 196 MeV. All the cross sections have been calculated with 
the phenomenological s.p. bound state wave functions of Ref.~\cite{ES}, which 
do not include correlations. The results shown in the left 
panels have been obtained without a correlation function, while the cross
sections in the right panels include the correlation function of 
Ref.~\cite{GD}, extracted form a recent application of Green's function 
techniques to the calculation of relative two-nucleon wave functions in nuclear 
matter. Correlation 
effects are small at 60 MeV, where they produce only a slight enhancement of 
the cross section that anyhow improves the agreement with data. At 196 MeV
SRC produce a strong enhancement of the contribution of the one-body current 
for nucleon emission angles above $70^{\rm o}$, in the region where high values 
of the recoil momentum are probed and the cross section is dominated by 
two-body currents. An opposite effect is given by the correlation function on 
the two-body current. The contributions of the $\Delta$ and seagull currents 
are significantly reduced by SRC at large angles, while correlation effects are 
not very important on the pion-in-flight term. In the final result the effect of
SRC is overwhelmed by the contribution of two-body currents, as it was found 
also in Ref.~\cite{Co}. The effect of SRC is anyhow non negligible and improves 
the agreement with data. 

The behaviour of the separate contributions of the different terms of the 
nuclear current is also shown in Fig.~\ref{fig:fig6}. At 
$E_{\gamma}= 60$ MeV the one-body current gives a large part of the final 
result and is already able to correctly reproduce the shape of the angular
distribution. The seagull current is also important, in particular at large 
scattering angles. The sum of the two terms produces, as in Fig.~\ref{fig:fig4}, 
a significant enhancement of the cross section. The pure contribution 
of  pion-in-flight is much smaller than those of the one-body and seagull 
currents, but it produces a strong destructive interference with seagull. The 
$\Delta$ current gives only a negligible contribution. At 196 MeV the one-body 
current gives only a small fraction of the experimental cross section. The 
final result is dominated by two-body currents, mainly by the isobar current 
at small proton emission angles and by the pion-in-flight term at large angles. 
The seagull current is not very important. At high values of the recoil 
momentum sizable and different contributions are given by SRC on the various 
components of the current. However, the combined effect does not appear large 
in the final result and is overwhelmed by MEC.

Our results for the $\left(\gamma,p\right)$ reactions are qualitatively similar 
to the results found in Ref.~\cite{Co}. The quantitative numerical differences 
are compatible with the different theoretical ingredients used in the two 
calculations, as the  $\left(\gamma,p\right)$ cross sections are very sensitive 
to details of the model. A
different choice of these ingredients might change the numerical results, but 
should not change the main conclusions of our work.


\section{Summary and conclusions}
\label{con}

The role of MEC in exclusive $\left(e,e^{\prime}p\right)$ and 
$\left(\gamma,p\right)$ reactions has been studied in the frame of a
nonrelativistic DKO model with final state interactions. The nuclear current is
the sum of a one-body part, including the convective and the spin terms, and of
a two-body part, including the contribution of all the diagrams with one-pion
exchange, namely seagull, pion-in-flight and the diagrams with intermediate
$\Delta$-isobar configurations. 
The direct contribution of the one-body current corresponds to the DKO mechanism
and gives the DWIA. Final state interactions are treated with a phenomenological
and spin dependent optical potential. Correlations can be included in the s.p.
wave functions or with a central correlation function.

Numerical results have been presented in different kinematics for electron- and
photon-induced reactions from $^{16}$O, in order to allow a comparison  with
previous calculation and to check the consistency of the description of the two
reactions also in comparison with data.

The role of two-body currents on the 
$^{16}$O$\left(e,e^{\prime}p\right)^{15}$N cross section is in general rather 
small. This result confirms once more the validity of the DWIA approach for 
this reaction. MEC give in general a small contribution also on the structure
functions. The effects are of about the same order and relevance as in the
analysis of Ref.~\cite{ALC}, but a different behaviour is found in our model on
the interference longitudinal-transverse structure function $f_{01}$ for the 
transitions to the  $1/2^{-}$ ground state and to first $3/2^{-}$ excited state 
of $^{15}$N. For the $1/2^{-}$ state the $\Delta$ current cancels the effect 
produced by the seagull term and the contribution of MEC is negligible. For the 
$3/2^{-}$ state the contributions of the two terms sum up and produce a sizable 
effect in the final result. A different behaviour on $f_{01}$ for the two states 
was found also in Ref.~\cite{Sluys}, where, however, the contribution of MEC, 
and in particular of the $\Delta$ current, is much larger than in our model. 

Results have been presented for the $^{16}$O$\left(\gamma,p\right)^{15}$N cross
section at $E_\gamma = 60$, 100 and 196 MeV. At all the considered photon 
energies MEC give an important contribution which substantially improves
the agreement with data. At $E_\gamma = 60$ MeV the DKO mechanism represents a
large part of the experimental cross section, but underestimates the size of the
data. The role of the one-body current on the cross section becomes less 
important increasing the photon energy and the proton emission angle. A 
significant enhancement is produced by the seagull current. At 60 MeV the 
seagull term gives the main contribution of the two-body currents, but 
overestimates their final effect. A strong destructive interference is obtained 
when the pion-in-flight term is added. The contribution of the seagull current 
decreases increasing the photon energy. Important contributions are given at 
higher energies by the pion-in-flight, in particular at large angles, and at 
196 MeV also by the $\Delta$ current. 

Significant and different effects on the various terms of the nuclear current 
are given by SRC at high values of the recoil momentum. The combined effect, 
however, does not appear large in the final result and is in general overwhelmed
by MEC. 

When all the one-pion exchange diagrams are included in the model, calculations
with consistent theoretical ingredients, i.e. bound state wave functions, 
optical potentials and the same spectroscopic factors, are able to give a 
reasonable and consistent description of the experimental cross sections of the 
$^{16}$O$\left(e,e^{\prime}p\right)^{15}$N and 
$^{16}$O$\left(\gamma,p\right)^{15}$N  reactions for recoil momentum values
up to $\sim 500-600$ MeV/$c$. For higher momentum values other effects might 
come into play and also a more careful treatment of correlations would 
presumably deserve further investigation.

\acknowledgments
We thank A.N. Antonov and M.K. Gaidarov for providing us the overlap functions
used in the calculations and G. Co' for useful discussions.

\end{document}